\documentclass[twocolumn,english,aps,prd,reprint,floatfix,notitlepage,footinbib,preprintnumbers,superscriptaddress,longbibliography]{revtex4-1}
\pdfoutput=1
\usepackage{lmodern}

\usepackage[T1]{fontenc}
\usepackage[latin9]{inputenc}
\usepackage{geometry}
\usepackage{subfigure,lmodern, amsmath,amssymb, graphicx, pifont, adjustbox, bm, xcolor}
\usepackage{amsfonts}
\usepackage{enumitem}
\usepackage{comment}
\usepackage{mathtools}
\usepackage{float}
\usepackage{slashed}
\usepackage{ragged2e}
\usepackage{array}
\usepackage{microtype}
\usepackage{soul}

\usepackage{nameref}

\usepackage{hhline}

\makeatletter\g@addto@macro\bfseries{\boldmath}\makeatother

\makeatletter\newcommand{\labeltext}[2]{%
  \def\@currentlabel{#1}%
  \label{#2}%
}
\makeatother

\usepackage{stackengine}
\usepackage{esint}
\usepackage[unicode=true,pdfusetitle,
 bookmarks=true,bookmarksnumbered=false,bookmarksopen=false,
 breaklinks=false,pdfborder={0 0 1},backref=false,colorlinks=true]
 {hyperref}
\hypersetup{
 pdfauthor={Clifford Cheung, Jaehoon Jeong, Pyungwon Ko, Alex Pomarol, Grant N. Remmen, Francesco Sciotti},
 citecolor=black,linkcolor=black,urlcolor=black}

\newcommand{\appendixref}[1]{\hyperref[#1]{appendix~\ref{#1}}}
\def\equationautorefname~#1\null{eq.\,(#1)\null}
\usepackage{breakurl}
\usepackage[hang,flushmargin]{footmisc} 
\allowdisplaybreaks
\makeatletter

\usepackage{etoolbox}
\apptocmd{\thebibliography}{\justifying\setlength{\leftskip}{7.4mm}}{}{} 
 
 \usepackage{relsize}
\usepackage{babel}

\usepackage{bbm}

\makeatletter
\def\simgt{\mathrel{\lower2.5pt\vbox{\lineskip=0pt\baselineskip=0pt
           \hbox{$>$}\hbox{$\sim$}}}}
\def\simlt{\mathrel{\lower2.5pt\vbox{\lineskip=0pt\baselineskip=0pt
           \hbox{$<$}\hbox{$\sim$}}}}
\makeatother

\usepackage{changepage}

\newcommand{\mysec}[1]{\noindent {\bf #1.}---}
\newcommand{\mysubsec}[1]{\noindent {\it #1.}---}

\newcolumntype{P}[1]{>{\centering\arraybackslash}p{#1}}

\usepackage{scalerel}

\usepackage{tikz}
\usetikzlibrary{calc}

\begin{document}

\preprint{KIAS-Q26019}

\title{Partial Waves for Multipositivity}

\author{Jaehoon Jeong}
\affiliation{\scalebox{1}{Quantum Universe Center, KIAS, Seoul 02455, Korea}}
    
\begin{abstract} 

\noindent 
We develop a partial-wave formalism providing the missing spin-resolved framework for multipositivity among massless planar amplitudes. We present a systematic procedure for constructing kinematic charts realizing the complex-forward limit. Applying it to obtain the four-, five-, and six-point charts, we show how to derive the one-to-two and one-to-three wavefunctions and construct the associated partial waves. The resulting formalism places higher-point multipositivity on the same footing as conventional partial-wave analyses of two-to-two scattering.

\noindent 
\end{abstract}
\maketitle 

\mysec{Introduction}
Recent years have witnessed remarkable progress in deriving model-independent constraints on effective field theories (EFTs) solely from general principles such as unitarity and causality~\cite{Adams:2006sv}, leading to powerful {\it positivity bounds} on Wilson coefficients of two-to-two scattering amplitudes~\cite{Bellazzini:2020cot,Arkani-Hamed:2020blm,Caron-Huot:2020cmc} with applications ranging from the Standard Model to quantum gravity~\cite{Albert:2022oes, Fernandez:2022kzi, Albert:2023jtd, Ma:2023vgc, Albert:2023seb, Li:2023qzs, Dong:2024omo, Dong:2025dpy,Nicolis:2009qm,Tolley:2020gtv,Pham:1985cr,Ananthanarayan:1994hf,Pennington:1994kc,Comellas:1995hq,Dita:1998mh,Manohar:2008tc,Mateu:2008gv,Jenkins:2006ia,Dvali:2012zc,completeness,Arkani-Hamed:2021ajd,Cheung:2018cwt,Cheung:2019cwi,EliasMiro:2022xaa,Caron-Huot:2021rmr,Bellazzini:2025shd,Bellazzini:2015cra,Cheung:2016wjt,Camanho:2014apa,Gruzinov:2006ie,Arkani-Hamed:2021ajd,Cheung:2018cwt,Cheung:2019cwi,Cheung:2014ega,Bellazzini:2019xts,Caron-Huot:2022ugt,Caron-Huot:2022jli,Cheung:2016yqr,deRham:2017xox,Bellazzini:2023nqj,Bern:2021ppb,Berman:2023jys,Freytsis:2022aho,Remmen:2019cyz,Remmen:2020vts,Remmen:2020uze,Bellazzini:2016xrt,Remmen:2022orj,Remmen:2024hry,Bi:2019phv,Zhang:2018shp,Berman:2024wyt,Remmen:2021zmc,Bachu:2022gof,Huang:2022mdb,Cheung:2022mkw,Cheung:2023adk,Cheung:2023uwn,Haring:2023zwu,Haring:2024wyz,Arkani-Hamed:2022gsa,Bhardwaj:2024klc,Aoude:2024xpx,SFAN,Guerrieri:2021ivu,Albert:2024yap,Berman:2024eid,Elvang:2026pmc,Cheung:2024uhn,Cheung:2024obl,Hillman:2024ouy,Caron-Huot:2016icg,Bellazzini:2025bay,Huang:2025icl,Beadle:2025cdx,Beadle:2024hqg,Pasiecznik:2025eqc,Caron-Huot:2024lbf,He:2023lyy,Cordova:2019lot,Guerrieri:2018uew,Guerrieri:2020bto,EliasMiro:2026kww,EliasMiro:2023fqi,EliasMiro:2022xaa,Eckner:2024ggx,Eckner:2024pqt,Cheung:2026dng,Albert:2026xyz,Chen:2026uyr,deRham:2026lvc,Shao:2026akl,Huang:2026ute,Cassem:2026aib,Xu:2026kix,Arkani-Hamed:2023jwn,Basile:2026gnd,Chandrasekaran:2018qmx,Berman:2025owb,Bresciani:2025toe,Cheung:2025nhw,Cheung:2026lpv,Saha:2026ftv}. Among these developments, partial-wave decomposition has played one of the central roles by providing a direct physical interpretation of the underlying UV spin structure.

Building on this success, recent efforts have begun to extend these ideas to higher multiplicity~\cite{Cheung:2025nhw,Arkani-Hamed:2023jwn,Bresciani:2025toe,Chandrasekaran:2018qmx,Berman:2025owb,Basile:2026gnd,Cheung:2026lpv,Saha:2026ftv}. In particular, Ref.~\cite{Cheung:2025nhw} uncovered qualitatively new positivity structures in higher-point massless planar amplitudes, known as {\it multipositivity}. A striking feature is the emergence of positivity relations between amplitudes with different numbers of external particles, with no counterpart in two-to-two scattering. Multipositivity has already found its first phenomenological application, leading to new constraints on the chiral Lagrangian~\cite{Cheung:2026lpv}.

A central ingredient in deriving multipositivity is the {\it complex-forward limit}~\cite{Cheung:2025nhw},
the complex extension of physical forward scattering.
In this limit, the cut structures entering dispersion relations can be
naturally organized in terms of positive quantities.
To fully identify the underlying UV states responsible for these positive cut
structures, the residues must be resolved into definite UV spin sectors.

Although general partial-wave formalisms exist for arbitrary multiplicity~\cite{Saha:2026ftv,Choi:2020mge,Jacob:1959at}, a natural approach would be to start from such a formalism and realize the complex-forward kinematics within that framework. In practice, however, this rapidly becomes unwieldy for higher-point multipositivity. Instead, we identify the kinematic conditions required by multipositivity and develop a systematic construction of momentum charts that automatically satisfy the Gram determinant constraints.

With this method, we parametrize four-, five-, and six-point momenta. On these kinematic charts, we obtain the one-to-two and one-to-three wavefunctions and construct the partial waves for four-, five-, and six-point amplitudes. The resulting formalism establishes the spin-resolved framework for multipositivity in massless planar amplitudes.

\medskip
\mysec{Kinematic Conditions for Multipositivity}In this work, we represent four-momenta by two-by-two matrices,
\begin{gather}
p_{i\alpha\dot \alpha}=\sigma^{\mu}_{\alpha\dot \alpha}p_{i\mu},
\end{gather}
where $\sigma^\mu$ denotes the Pauli matrices. This will be expanded by the matrix basis,
\begin{gather}
\ell_+= 
\begin{pmatrix}
1&0
\\
0&0
\end{pmatrix},\quad
\ell_-= 
\begin{pmatrix}
0&0
\\
0&1
\end{pmatrix},
\\
n_+= 
\begin{pmatrix}
0&1
\\
0&0
\end{pmatrix},\quad 
n_-= 
\begin{pmatrix}
0&0
\\
1&0
\end{pmatrix},
\end{gather}
with $\det(\ell_+ +\ell_-)=-\det(n_+ +n_-)=1$ and all other contractions vanishing~\footnote{These matrices form a tetrad basis in the four-momentum space~\cite{Cheung:2026lpv}.}.

Consider the $(n+m)$-point kinematics with the external momenta divided into two sets,
$(p_{1,\ldots,n}\,|\,p_{n+1,\ldots,n+m})$\footnote{Throughout this work, we use the shorthand notation $p_{1,\ldots,N}\equiv(p_1,\ldots,p_N)$ for arbitrary $N$.}.
The momentum conservation and on-shell conditions are
\begin{gather}
p_1+...+p_{n}=-(p_{n+1}+...+p_{n+m}),
\label{eq:mom_conserve}
\\
\det p_i=0,
\label{eq:on-shell}
\end{gather}
where Eq.~\eqref{eq:on-shell} states that each massless momentum is represented by a rank-one matrix. We define the factorization invariant and the other Mandelstam invariants by
\begin{gather}
s=(p_1+\cdots+p_n)^2=(p_{n+1}+\cdots+p_{n+m})^2,
\\
s_{i_1...i_l}=(p_{i_1}+...+p_{i_l})^2,
\end{gather}
in mostly-minus signature, where $i_1,\ldots,i_l$ are ordered so that the planar Mandelstams form a basis.

To derive multipositivity, we first examine the residue of an $(n+m)$-point amplitude at the heavy-state pole $s=m_k^2$, which factorizes into two lower-point amplitudes~\footnote{In the all-incoming convention, the conjugation of an $N$-point amplitude is
\(A^{(N)*}(p_{1,\ldots,N})=A^{(N)}(-p_{N,\ldots,1}^{\dagger})\)}:
\begin{align}
&R^{(n+m)}_k(p_{1},...,p_{n}|p_{n+1},...,p_{n+m})
\nonumber
\\
&=A^{(1+n)}_k(p_{1},...,p_{n})
A^{(1+m)*}_k(-p_{n+m}^{\dagger},...,-p_{n+1}^{\dagger}),
\label{eq:residue}
\end{align}
where we omitted the external helicities and left the sum over exchanged-state quantum numbers (e.g. helicity) implicit. 
\begin{figure*}[!t]
    \centering
    \includegraphics[width=1\linewidth]{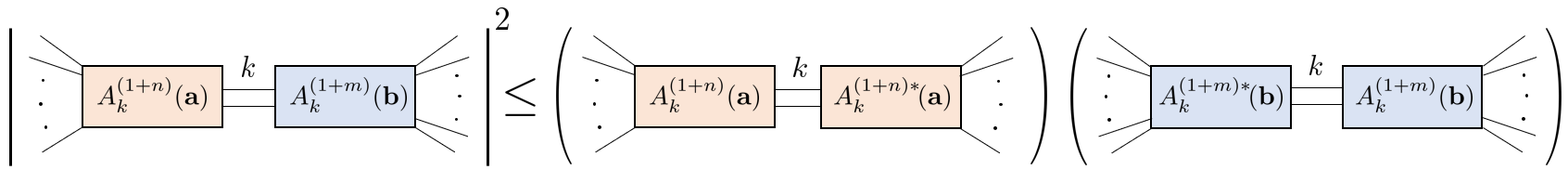}
    \caption{Diagrammatic illustration of Eq.~\eqref{eq:cauchy-schwarz}.}
    \label{fig:cauchy-schwarz}
\end{figure*}
Next, we parametrize the momenta as $p_i(s,\mathbf c)$ for $i=1,\ldots,n+m$, in terms of the factorization invariant $s$ and a set of remaining parameters $\mathbf c$. To construct other kinematics required for the Cauchy--Schwarz inequality, we define
\begin{align}
\bar{p}_i(s,{\bf c})\equiv p^\dagger_i(s^*,{\bf c}),
\end{align}
so that $\bar p_i=p_i^\dagger$ for real $s$. 

With this definition, the amplitudes of two combined kinematics
$(p_{1,\ldots,n}|-\bar p_{n,\ldots,1})$ and
$(p_{n+1,\ldots,n+m}|-\bar p_{n+m,\ldots,n+1})$
reproduce the residues $|A_k^{(1+n)}|^2$ and $|A_k^{(1+m)}|^2$, respectively. Here, the definition $\bar p_i$ automatically satisfies the on-shell condition, while momentum conservation does not. Thus, we require the first kinematic condition for multipositivity:
\vspace{0.4cm}

\noindent
\phantomsection\label{cond:1}%
\begin{tabular*}{\linewidth}{@{}l@{\hspace{2em}}c@{}}
$\langle1\rangle$ &\;\;
$\begin{aligned}
p_1+\cdots+p_n
&=\bar p_n+\cdots+\bar p_1,\\
p_{n+1}+\cdots+p_{n+m}
&=\bar p_{n+m}+\cdots+\bar p_{n+1},
\end{aligned}$
\end{tabular*}
\vspace{0.2cm}

\noindent
leading the two combined kinematics to realize the complex-forward limit for real $s\ge 0$, respectively.

To express the two lower-point amplitudes independently, we split the momentum parametrization as
\begin{align}
p_{1,\ldots,n}(s, {\bf a})
,\quad
p_{n+1,\ldots,n+m}(s,{\mathbf b}),
\label{eq:momenta}
\end{align}
such that the two sets of parameters $\bf a$ and $\bf b$ make the Mandelstams regular in their vanishing limit. 

Then, collecting the residues of $(n+m)$-, $2n$-, and $2m$-point amplitudes in our kinematic charts, we get the Cauchy--Schwarz inequality:
\begin{align}
&\Big|\sum_k 
\Big(A_k^{(1+n)}({\bf a})\Big)
\Big(A_k^{(1+m)}({\bf b})\Big)\Big|^2
\nonumber
\\
&\leq 
\Big(\sum_k \Big| A_k^{(1+n)}({\bf a})\Big|^2
\Big)
\Big(
\sum_k
\Big| A_k^{(1+m)}({\bf b})\Big|^2
\Big).
\label{eq:cauchy-schwarz}
\end{align}
To directly apply Eq.~\eqref{eq:cauchy-schwarz} to the analysis of
multipositivity for massless planar amplitudes, we require the second kinematic condition:

\vspace{0.2cm}
\noindent
\phantomsection\label{cond:2}%
\begin{minipage}[c]{0.08\linewidth}
  \raggedright
  $\langle 2\rangle$
\end{minipage}%
\begin{minipage}[c]{0.90\linewidth}
  The factorization invariant $s$ is independent of all other
  kinematic invariants in the $(n+m)$-, $2n$-, and $2m$-point
  amplitudes.
\end{minipage}

\medskip
\mysec{Momentum parametrizations}Based on the two conditions~\hyperref[cond:1]{$\langle 1\rangle$} and~\hyperref[cond:2]{$\langle 2\rangle$}, we set up the logic of the momentum parametrization. As the first nontrivial example, consider $(p_{1,2}|p_{3,4,5})$. To satisfy the condition~\hyperref[cond:1]{$\langle 1\rangle$}, we take the parametrization in the center of mass (c.m.) frame: 
\begin{gather}
p_1+p_2=-p_{3}-p_{4}-p_{5}=\sqrt{s}\mathbf 1.
\end{gather}
Next, we construct a two-momentum block such that the kinematics $(p_{1,2}|-\bar{p}_{2,1})$ satisfies the condition~\hyperref[cond:2]{$\langle 2\rangle$}. We then extend it to a three-momentum block so that the $(p_{1,2}|p_{3,4,5})$ kinematics satisfies the condition~\hyperref[cond:2]{$\langle 2\rangle$}, and finally verify the condition~\hyperref[cond:2]{$\langle 2\rangle$} for the kinematics $(-\bar{p}_{5,4,3}|p_{3,4,5})$.

\medskip
\mysubsec{$(p_{1,2}|p_{3,4})$ Kinematics} We begin with a reference configuration involving
only the $\ell_\pm$ matrices:
\begin{align}
q_1=\sqrt{s}\,\ell_+,\quad
q_2=\sqrt{s}\,\ell_- .
\end{align}
Its combined kinematics $(q_{1,2}|-\bar{q}_{2,1})$ realizes
the physical forward limit for real $s\ge 0$, where the mixed Mandelstam vanishes,
$s_{23}=\det(q_2-q_2^\dagger)=0$.

To turn on this Mandelstam under the conditions in
Eqs.~\eqref{eq:mom_conserve} and~\eqref{eq:on-shell}, we shift
$p_{1,2}= q_{1,2}\pm a n_+$~\footnote{A shift with the $\ell_\pm$ matrices
cannot turn on the mixed Mandelstam because of diagonality.},
which follows the shift introduced in Ref.~\cite{Cheung:2025nhw}, itself inspired by on-shell recursion relations~\cite{Britto:2005fq}. This gives
$s_{23}=\det(a n_+-a^* n_-)=|a|^2$ in the chart $(p_{1,2}|-\bar{p}_{2,1})$. 

For the other side, to satisfy the condition~\hyperref[cond:2]{$\langle 2\rangle$}, we choose $p_{3,4}=-p^T_{2,1}(a\to b)$, which leads to the mixed Mandelstam $s_{23}=\det(p_2+p_3)=ab$. This choice automatically makes the combined kinematics $(-\bar{p}_{4,3}|p_{3,4})$ obey the condition~\hyperref[cond:2]{$\langle 2\rangle$}. Thus, we have 
\begin{gather}
\begin{aligned}
p_1&=
\begin{pmatrix}
\sqrt{s}&a
\\
0&0
\end{pmatrix}
\\
p_2&=
\begin{pmatrix}
0&-a
\\
0&\sqrt{s}
\end{pmatrix}
\end{aligned}
\left|\;
\begin{aligned}
p_4&=-
\begin{pmatrix}
\sqrt{s}&0
\\
b&0
\end{pmatrix}
\\
p_3&=-
\begin{pmatrix}
0&0
\\
-b&\sqrt{s}
\end{pmatrix}
\end{aligned}.
\right.
\label{eq:2-to-2_momenta}
\end{gather}

\medskip
\mysubsec{$(p_{1,2}|p_{3,4,5})$ Kinematics}Previously, we already found the two-momentum chart $p_{1,2}$ in Eq.~\eqref{eq:2-to-2_momenta} realizing the complex-forward completion. We thus move on to the three-momentum side. Let us start from the reference momenta 
\begin{align}
q_5=-\sqrt{s}\ell_+,\;\; q_4=-\frac{s_{45}}{\sqrt{s}}\ell_-, \;\; q_3=-\frac{s-s_{45}}{\sqrt{s}}\ell_-,
\end{align}
which yield the vainishing mixed Mandelstams $s_{51}=s_{23}=0$. Here, we introduced the internal invariant $s_{45}$ used as a parameter of the three-momentum side. 

To turn on another internal Mandelstam, we shift 
\begin{gather}
q_{5}'=-
\begin{pmatrix}
\sqrt{s}&-b_2
\\
b_1&0
\end{pmatrix},
\nonumber
\\
q_{4}'=-
\begin{pmatrix}
0&b_2
\\
0&\frac{s_{45}}{\sqrt{s}}
\end{pmatrix},
\;\;
q_{3}'=-
\begin{pmatrix}
0&0
\\
-b_1&\frac{s-s_{45}}{\sqrt{s}}
\end{pmatrix},
\label{eq:3p_shifts}
\end{gather}
leading to $s_{34}=b_1b_2$. The shifted momenta, then, automatically turn on $s_{23}=ab_1$. However, $q_5'$ violates the on-shell condition.
It leads us to shift
$q_{5,4}'\to q_{5,4}'\pm b_1b_2\ell_-/\sqrt{s}$. We then have
\begin{gather}
\begin{aligned}
p_1&=
\begin{pmatrix}
\sqrt{s}&a
\\
0&0
\end{pmatrix}
\\
p_2&=
\begin{pmatrix}
0&-a
\\
0&\sqrt{s}
\end{pmatrix}
\end{aligned}
\left|\;
\begin{aligned}
p_{5}&=-
\begin{pmatrix}
\sqrt{s}&-b_2
\\
b_1&-\frac{b_1b_2}{\sqrt{s}}
\end{pmatrix}
\\
p_{4}&=-
\begin{pmatrix}
0&b_2
\\
0&\frac{s_{45}+b_1b_2}{\sqrt{s}}
\end{pmatrix}
\\
p_{3}&=-
\begin{pmatrix}
0&0
\\
-b_1&\frac{s-s_{45}}{\sqrt{s}}
\end{pmatrix}
\end{aligned}
\right..
\label{eq:2-to-3_momenta}
\end{gather}
The planar Mandelstams $\{s_{12},s_{23},s_{34},s_{45},s_{51}\}$ in this chart are
\begin{align}
\big\{s, a b_1, b_1b_2, s_{45}, (a+b_2)b_1\big\},
\end{align}
where the last entry satisfies $s_{51}=s_{23}+s_{34}$.

Other charts can be obtained by setting $s_{34}=0$ or $s_{45}=0$
(see Ref.~\cite{Cheung:2025nhw} for the $s_{45}=0$ chart).
Such charts are useful for theories without massless three-point amplitudes~\cite{Cheung:2015ota,Padilla:2016mno,Low:2019ynd}; otherwise, higher-point amplitudes become singular when a two-particle Mandelstam vanishes.
We therefore discuss the charts with all Mandelstams nonzero in this work.

\medskip
\mysubsec{$(p_{1,2,3}|p_{4,5,6})$ Kinematics}In the derivation of the
kinematics $(p_{1,2}|p_{3,4,5})$, we did not check the condition~\hyperref[cond:2]{$\langle 2\rangle$} for the combined kinematics $(-p_{5,4,3}^\dagger|p_{3,4,5})$, since it can be checked directly in the
$(p_{1,2,3}|p_{4,5,6})$ kinematics. Using Eq.~\eqref{eq:2-to-3_momenta}, we have
\begin{gather}
\begin{aligned}
p_{1}&=
\begin{pmatrix}
\sqrt{s}&a_1
\\
-a_2&-\frac{a_1a_2}{\sqrt{s}}
\end{pmatrix}
\\
p_{2}&=
\begin{pmatrix}
0&0
\\
a_2&\frac{s_{12}+a_1a_2}{\sqrt{s}}
\end{pmatrix}
\\
p_{3}&=
\begin{pmatrix}
0&-a_1
\\
0&\frac{s-s_{12}}{\sqrt{s}}
\end{pmatrix}
\end{aligned}
\left|\;
\begin{aligned}
p_{6}&=-
\begin{pmatrix}
\sqrt{s}&-b_2
\\
b_1&-\frac{b_1b_2}{\sqrt{s}}
\end{pmatrix}
\\
p_{5}&=-
\begin{pmatrix}
0&b_2
\\
0&\frac{s_{56}+b_1b_2}{\sqrt{s}}
\end{pmatrix}
\\
p_{4}&=-
\begin{pmatrix}
0&0
\\
-b_1&\frac{s-s_{56}}{\sqrt{s}}
\end{pmatrix}
\end{aligned}
\right.,
\label{eq:3-to-3_momenta}
\end{gather}
where we used $p_{6,5,4}=-p^T_{1,2,3}(a_{1,2}\to b_{1,2})$ with $s_{12} \to s_{56}$. The planar Mandelstams are then
\begin{align}
&\{s_{12},s_{23},s_{34},s_{45},s_{56},s_{61}\}
\nonumber
\\
&=\{s_{12},a_1a_2,a_1b_1,b_1b_2,s_{56},(a_1+b_2)(b_1+a_2)\},
\\
&\{s_{123},s_{234},s_{345}\}
\nonumber
\\
&=\{s,(b_1+a_2)a_1,(a_1+b_2)b_1\},
\end{align}
where $s_{61}$, $s_{234}$, and $s_{345}$ are not independent but are fixed by $s_{23}$, $s_{34}$, and $s_{45}$. In this way, one can continue the construction to the general chart $(p_{1,\ldots,n}|p_{n+1,\ldots,n+m})$.

\medskip

\mysec{One-to-Many Wavefunctions} For the $(n+m)$-point amplitude, the PW can be written in terms of the angular momentum $J$ as
\begin{align}
\begin{aligned}
&\!\!\!A^{(n+m)}(p_1^{h_1},\!...,p_{n}^{h_n}|p_{n+1}^{h_{n+1}},\!...,p_{n+m}^{h_{n+m}})
=\overline{\sum}\,H^J_{\tau|\tau'} \!\!
\\
&\!\!\!\times
\langle JM,\tau'|\!-\!\bar{p}_{n+m}^{\,h_{n+m}},\!...,-\bar{p}_{n+1}^{\,h_{n+1}}\rangle^*
\langle JM,\tau|p_{1}^{h_{1}},\!...,p_{n}^{h_{n}}\rangle,\!\!
\end{aligned}
\label{eq:general_PW}
\end{align}
where $p_i^{h_i}$ denotes the momentum and helicity, $\tau,\tau'$ are sets of quantum numbers, and $\overline{\sum}$ sums over $J,M,\tau,\tau'$. Here, we omitted the helicity dependence in $H^J$ and $|JM,\tau\rangle$. Although we also omitted the Mandelstam dependence in $H^J$ for now, it will be specified later.

For the single factorization invariant $s$, the residue of
Eq.~\eqref{eq:general_PW} at a heavy-state pole $s=m_k^2$ factorizes the
coefficient matrix as $H^J_{\tau|\tau'}\to g^{J}_{k,\tau}g^{J*}_{k,\tau'}$. The two couplings carry the quantum numbers of the two wavefunction blocks, $\langle JM,\tau|p_{1,\ldots n}\rangle$ and $\langle JM,\tau'|\!-\!\bar{p}_{n+m,\ldots n+1}\rangle^*$, respectively.

Thus, the main task in constructing the PWs is to find the explicit form of the
wavefunctions $\langle JM,\tau|p_{1}^{h_{1}},\!...,p_{n}^{h_{n}}\rangle$ as functions of the momentum parameters for the angular quantum numbers $J,M,\tau$ and $h_i$. By
identifying these wavefunctions on our kinematic charts, we complete the PW expansions.

\medskip
\mysubsec{One-to-Two Wavefunction} We first note that a Lorentz transformation $p \to \Lambda p\Lambda^\dagger$ is represented on momentum eigenstates by $U[\Lambda]|p\rangle$ on the
physical slice. In our kinematic charts, we extend the transformation to
holomorphically complexified rotations and boosts acting as $RpR^{-1}$ and
$LpL$, respectively.

We introduce the rotation generators acting on the two-by-two momentum matrices:
\begin{gather}
J_{+}=
\begin{pmatrix}
0&1
\\
0&0
\end{pmatrix}
,\;\;
J_{-}=
\begin{pmatrix}
0&0
\\
1&0
\end{pmatrix}
,\;\; 
J_{0}=\frac12
\begin{pmatrix}
1&0
\\
0&-1
\end{pmatrix},
\end{gather}
satisfying the commutators,
\begin{gather}
[J_+,\ell_{\pm}]=\mp n_+,\quad [J_-,\ell_{\pm}]=\pm n_-,
\label{eq:J_pm_comm}
\\
[J_0,\ell_\pm]=0,\quad [J_0,n_{\pm}]=\pm n_\pm.
\label{eq:J_0_comm}
\end{gather}
We also define the complexified rotations
$R_{\pm,0}(z)=e^{-z J_{\pm,0}}$ for a parameter $z$. Their matrix forms are
\begin{gather}
R_{+}(z)= 
\begin{pmatrix}
1&-z
\\
0&1
\end{pmatrix}
,\quad
R_{-}(z)= 
\begin{pmatrix}
1&0
\\
-z &1
\end{pmatrix},
\nonumber
\\
R_{0}(z)=
\begin{pmatrix}
e^{-z/2}&0
\\
0&e^{z/2}
\end{pmatrix}.
\end{gather}
We express a generic complexified rotation in a Gauss-decomposed form,
\begin{gather}
\mathcal{R}(\vec z)=R_{+}(z_1)R_{-}(z_2)R_{0}(z_3),
\end{gather}
with parameters $\vec z=(z_1,z_2,z_3)$.

Recall that the two momenta $(p_1,p_2)$ in Eq.~\eqref{eq:2-to-2_momenta} were obtained by shifting $(\sqrt{s}\ell_+,\sqrt{s}\ell_-)$. This shift is realized by the complexified rotation:
\begin{align}
&(p_1,p_2)=R_+\bigg(\frac{b}{\sqrt{s}}\bigg)(\sqrt{s}\ell_+,\sqrt{s}\ell_-)
R_+^{-1}\bigg(\frac{b}{\sqrt{s}}\bigg).
\end{align}
Thus, the two-particle state with momenta $(p_1,p_2)$ can be written as
\begin{gather}
|p_1^{h_1},p_2^{h_2}\rangle=U\!\left[R_+\!\left(\frac{a}{\sqrt{s}}\right)\right]|\sqrt{s}\ell_+^{h_1},\sqrt{s}\ell_-^{h_2}\rangle.
\end{gather}
Eq.~\eqref{eq:J_0_comm} implies that
$|\sqrt{s}\ell_+^{h_4},\sqrt{s}\ell_-^{h_3}\rangle$ has $J_0$ eigenvalue
$h_{43}=h_4-h_3$. Thus, in a fixed-$J$ partial wave, we can project it onto
$|Jh_{43}\rangle$. This gives
\begin{gather}
\langle JM|p_4^{h_{4}}\!,p_3^{h_{3}}\rangle \propto
\langle JM|U\!\left[R_+\!\left(\frac{b}{\sqrt{s}}\right)\right]|J h_{43}\rangle.
\label{eq:2p_wave}
\end{gather}

\medskip
\mysubsec{One-to-Three Wavefunctions} For the three-particle state, we use the momenta $(p_1,p_2,p_3)$ in
Eq.~\eqref{eq:3-to-3_momenta}. We choose a reference frame in which the (1,2) subsystem momentum $K_{12}=k_1+k_2$ and $k_3$ take a
simple diagonal form. This is obtained by the transformation
\begin{align}
|p_1^{h_1}\!,p_2^{h_2}\!,p_3^{h_3}\rangle 
=&U\!\left[
R_+\!\left(
\frac{a_1\sqrt{s}}{s-s_{12}}
\right)
\right]
|k_1^{h_1}\!,k_2^{h_2}\!,k_3^{h_3}\rangle,
\label{eq:p123_ref}
\end{align}
where
\begin{align}
&k_1
=
\begin{pmatrix}
\frac{\sqrt{s}(s-s_{12}-a_1a_2)}{s-s_{12}}
&
-\frac{a_1s_{12}(s-s_{12}-a_1a_2)}{(s-s_{12})^2}
\\[8pt]
-a_2
&
\frac{a_1a_2s_{12}}{\sqrt{s}(s-s_{12})}
\end{pmatrix},
\nonumber
\\
&k_2
=
\begin{pmatrix}
\frac{a_1a_2\sqrt{s}}{s-s_{12}}
&
\frac{a_1s_{12}(s-s_{12}-a_1a_2)}{(s-s_{12})^2}
\\[8pt]
a_2
&
\frac{s_{12}(s-s_{12}-a_1a_2)}{\sqrt{s}(s-s_{12})}
\end{pmatrix}
,\;
k_3
=
\begin{pmatrix}
0&0
\\
0&\frac{s-s_{12}}{\sqrt{s}}
\end{pmatrix}.
\end{align}

The reference state can be decomposed by first resolving the $K_{12}$
subsystem into angular-momentum states:
\begin{align}
&|k_1^{h_1}\!,k_2^{h_2} \!,k_3^{h_3}\rangle
\nonumber
\\
&=\sum_{J_2,M_2} 
|K_{12};J_2M_2 \rangle \otimes |k_3^{h_3}\rangle 
\langle J_2M_2| \hat k_1^{h_1}\! , \hat k_2^{h_2}\rangle,
\label{eq:k_p123}
\end{align}
where $\hat k_1,\hat k_2$ denote the momenta in their c.m. frame, so that
$\hat K_{12}=\sqrt{s_{12}}\mathbf{1}$. This frame is related to the $K_{12}$ frame by $K_{12}=L_z\hat K_{12}L_z$, with the complexified Lorentz boost $L_z=\mbox{diag}[(s/s_{12})^{1/4},(s_{12}/s)^{1/4}]$.

The two-particle state in the $\hat{K}_{12}$ subsystem is
\begin{align}
|\hat k_1^{h_1},\hat k_2^{h_2}\rangle
&=
U\bigg[
\mathcal{R}\bigg(-\frac{a_1\sqrt{s_{12}}}{s-s_{12}},\frac{a_2}{\sqrt{s_{12}}},0\bigg)
\bigg]
\nonumber
\\
&\quad\times
|\sqrt{s_{12}}\ell_+^{h_1},\sqrt{s_{12}}\ell_-^{h_2}\rangle .
\end{align}
As in the one-to-two case,
$|\sqrt{s_{12}}\ell_+^{h_1},\sqrt{s_{12}}\ell_-^{h_2}\rangle$
has $J_0$ eigenvalue $h_{12}=h_1-h_2$. Therefore, in a fixed-$J_2$ partial wave, we project it onto $|J_2h_{12}\rangle$. Thus, the wavefunction $\langle J_2M_2 |\hat k_1^{h_1}\!,\hat k_2^{h_2}\!\rangle$ is proportional to
\begin{align}
\langle J_2M_2|U\bigg[
\mathcal{R}\bigg(-\frac{a_1\sqrt{s_{12}}}{s-s_{12}},\frac{a_2}{\sqrt{s_{12}}},0\bigg)
\bigg]|J_2 h_{12}\rangle.
\label{eq:k12}
\end{align}
\vspace{-1pt}

In Eq.~\eqref{eq:k_p123}, the momentum $k_3$ is proportional to
$\ell_-$, so $|k_3\!\!\!{}^{\,h_3}\rangle$ has $J_0$ eigenvalue $-h_3$.
Together with $|K_{12};J_2M_2\rangle$, whose $J_0$ eigenvalue is $M_2$,
the tensor product has total $J_0$ eigenvalue $M_2-h_3$. Therefore, in a
fixed-$J_1$ partial wave, we project it onto $|J_1,M_2-h_3\rangle$. Collecting Eqs.~\eqref{eq:p123_ref},~\eqref{eq:k_p123}, and~\eqref{eq:k12}, we have 
\begin{align}
&\langle J_1M_1;J_2M_2|p_1^{h_1} \!,p_2^{h_2} \!,p_3^{h_3}\rangle  
\nonumber
\\
&\propto\langle J_1M_1|U\bigg[R_+\!\left(
\frac{a_1\sqrt{s}}{s-s_{12}}
\right)\bigg]|J_1,M_2-h_3\rangle 
\nonumber
\\
&\quad \times
\langle J_2M_2|U\bigg[
\mathcal{R}\bigg(-\frac{a_1\sqrt{s_{12}}}{s-s_{12}},\frac{a_2}{\sqrt{s_{12}}},0\bigg)
\bigg] |J_2 h_{12}\rangle.
\label{eq:3p_wave}
\end{align}

\medskip
\mysubsec{Partial Waves for Multipositivity} Using the wavefunctions in Eqs.~\eqref{eq:2p_wave} and~\eqref{eq:3p_wave}, we can construct the PWs for the four-, five-, and six-point massless amplitudes~\footnote{Note that the PW construction itself is independent of planarity. Thus, nonplanar amplitudes can also be expanded in the PWs derived here.}. By defining the rotation matrix element $F^J_{MM'}(\vec z)\equiv \langle JM|U[\mathcal{R}(\vec z)]|JM'\rangle$ and $f^J_{MM'}(z)\equiv F^J_{MM'}(z,0,0)$, we have~\footnote{The coefficients in our PWs are the usual ones in the conventional PW 
expansion, defined in the angular-momentum basis. For example, for a four-point
amplitude one may write
$H^J(s)=\langle JM|\mathcal{M}(s)|JM\rangle$, with the normalization
factors between the momentum and angular-momentum bases absorbed into
$H^J$.}
\begin{widetext}
\begin{align}
&A^{(4)}_{12|34}
=\overline{\sum}\,
f^J_{Mh_{12}}\bigg(\frac{a}{\sqrt{s}}\bigg)
H^J(s) \,
f^J_{Mh_{43}}\bigg(\frac{b}{\sqrt{s}}\bigg),
\label{eq:4pt_PWs}
\\
&A^{(5)}_{12|345}
=\overline{\sum}\,
f^{J_1}_{M_1 h_{12}}\bigg(\frac{a}{\sqrt{s}}\bigg)
H^{J_1}_{|J_2M_2}(s,s_{45})\,
f^{J_1}_{M_1,M_2-h_3}\bigg(\frac{b_1\sqrt{s}}{s-s_{45}}\bigg)
F^{J_2}_{M_2 h_{54}}\bigg(
-\frac{b_1\sqrt{s_{45}}}{s-s_{45}},
\frac{b_2}{\sqrt{s_{45}}},
0
\bigg),
\label{eq:5pt_PWs}
\\
&A^{(6)}_{123|456}
=\overline{\sum}\,
f^{J_2}_{M_2,M_2-h_3}\bigg(\frac{a_1\sqrt{s}}{s-s_{12}}\bigg)
F^{J_1}_{M_1 h_{12}}\bigg(
-\frac{a_1\sqrt{s_{12}}}{s-s_{12}},
\frac{a_2}{\sqrt{s_{12}}},
0
\bigg)
H^{J_2}_{J_1M_1 |J_3M_3}(s_{12},s,s_{56})
\nonumber
\\
&\qquad\qquad\qquad \times
f^{J_2}_{M_2,M_3-h_4}\bigg(\frac{b_1\sqrt{s}}{s-s_{56}}\bigg)
F^{J_3}_{M_3 h_{65}}\bigg(
-\frac{b_1\sqrt{s_{56}}}{s-s_{56}},
\frac{b_2}{\sqrt{s_{56}}},
0
\bigg),
\label{eq:6pt_PWs}
\end{align}
\end{widetext}
where $\overline{\sum}$ denotes the sum over all angular momenta $J_i$ and magnetic quantum numbers $M_i$ appearing in each PW expansion. Since these PWs are constructed on kinematic charts satisfying the conditions~\hyperref[cond:1]{$\langle 1\rangle$} and~\hyperref[cond:2]{$\langle 2\rangle$}, they directly provide the spin-resolved ingredients required for the multipositivity bounds, $|4pt|^2\leq (4pt)(4pt)$, $|5pt|^2\leq (4pt)(6pt)$, and $|6pt|^2\leq (6pt)(6pt)$. As a first application of the formalism, Appendix applies Eq.~\eqref{eq:5pt_PWs} to the five-point Wess-Zumino-Witten amplitude~\cite{Wess:1971yu,Witten:1983tw}, demonstrating how low-energy EFT data are resolved into definite-spin UV information.

\newcommand{\J}{\mathcal{J}}

\medskip
\mysec{Discussion}We have developed a partial-wave formalism directly applicable to multipositivity among massless planar amplitudes. The key step was to identify the kinematic conditions required by multipositivity and to present a systematic procedure for constructing momentum charts. Applying this procedure, we constructed the four-, five-, and six-point momentum charts, derived the corresponding wavefunctions using complexified rotations, and constructed the partial waves. The resulting formalism establishes the spin-resolved framework for higher-point multipositivity, which places higher-point multipositivity on the same footing as conventional partial-wave analyses of two-to-two scattering.

Beyond the examples presented here, the same methodology applies directly to other kinematic charts with vanishing two-particle Mandelstam invariants, which can be particularly useful in theories without massless three-point amplitudes, such as chiral perturbation theory. More generally, it provides a systematic route toward constructing the $(n+m)$-, $2n$-, and $2m$-point kinematics required for higher-point multipositivity.

Several directions naturally follow from this work. One is to develop a systematic procedure for constructing the kinematics required for nonplanar multipositivity, together with the partial waves constructed on such kinematics. Another is to extend the present framework to massive kinematics and the corresponding partial waves. In particular, massive two-to-two partial waves based on kinematics realizing the complex-forward limit will be reported separately.

\medskip
\noindent {\it Acknowledgments:} 
We thank Alex Pomarol for valuable comments. J.J. has been supported by KIAS Individual Grants (QP090001) via the Quantum Universe Center at Korea Institute for Advanced Study.

\bibliographystyle{utphys-modified}
\bibliography{PW_multipos}

@article{Mateu:2008gv,
    author = "Mateu, Vicent",
    title = "{Universal Bounds for $SU(3)$ Low Energy Constants}",
    eprint = "0801.3627",
    archivePrefix = "arXiv",
    primaryClass = "hep-ph",
    reportNumber = "IFIC-08-02, FTUV-07-0124",
    doi = "10.1103/PhysRevD.77.094020",
    journal = "Phys. Rev. D",
    volume = "77",
    pages = "094020",
    year = "2008"
}

@article{Manohar:2008tc,
    author = "Manohar, Aneesh V. and Mateu, Vicent",
    title = "{Dispersion Relation Bounds for pi pi Scattering}",
    eprint = "0801.3222",
    archivePrefix = "arXiv",
    primaryClass = "hep-ph",
    reportNumber = "IFIC-08-01, FTUV-07-0121",
    doi = "10.1103/PhysRevD.77.094019",
    journal = "Phys. Rev. D",
    volume = "77",
    pages = "094019",
    year = "2008"
}

@article{Dita:1998mh,
    author = "Di\c{t}\u{a}, Petre",
    title = "{Positivity constraints on chiral perturbation theory pion pion scattering amplitudes}",
    eprint = "hep-ph/9809568",
    archivePrefix = "arXiv",
    doi = "10.1103/PhysRevD.59.094007",
    journal = "Phys. Rev. D",
    volume = "59",
    pages = "094007",
    year = "1999"
}

@article{Comellas:1995hq,
    author = "Comellas, Jordi and Latorre, Jose Ignacio and Taron, Josep",
    title = "{Constraints on chiral perturbation theory parameters from QCD inequalities}",
    eprint = "hep-ph/9507258",
    archivePrefix = "arXiv",
    reportNumber = "UB-ECM-PF-95-14",
    doi = "10.1016/0370-2693(95)01110-C",
    journal = "Phys. Lett. B",
    volume = "360",
    pages = "109",
    year = "1995"
}

@article{Cheung:2025nhw,
    author = "Cheung, Clifford and Remmen, Grant N.",
    title = "{Multipositivity bounds for scattering amplitudes}",
    eprint = "2505.05553",
    archivePrefix = "arXiv",
    primaryClass = "hep-th",
    reportNumber = "CALT-TH 2025-010",
    doi = "10.1103/wt4x-2149",
    journal = "Phys. Rev. D",
    volume = "112",
    pages = "016017",
    year = "2025"
}

@article{He:2023lyy,
    author = "He, Yifei and Kruczenski, Martin",
    title = "{Bootstrapping gauge theories}",
    eprint = "2309.12402",
    archivePrefix = "arXiv",
    primaryClass = "hep-th",
    doi = "10.1103/PhysRevLett.133.191601",
    journal = "Phys. Rev. Lett.",
    volume = "133",
    pages = "191601",
    year = "2024"
}

@article{Cordova:2019lot,
    author = "C{\'o}rdova, Luc{\'\i}a and He, Yifei and Kruczenski, Martin and Vieira, Pedro",
    title = "{The O(N) S-matrix Monolith}",
    eprint = "1909.06495",
    archivePrefix = "arXiv",
    primaryClass = "hep-th",
    doi = "10.1007/JHEP04(2020)142",
    journal = "JHEP",
    volume = "04",
    pages = "142",
    year = "2020"
}

@article{Guerrieri:2018uew,
    author = "Guerrieri, Andrea L. and Penedones, Joao and Vieira, Pedro",
    title = "{Bootstrapping QCD Using Pion Scattering Amplitudes}",
    eprint = "1810.12849",
    archivePrefix = "arXiv",
    primaryClass = "hep-th",
    doi = "10.1103/PhysRevLett.122.241604",
    journal = "Phys. Rev. Lett.",
    volume = "122",
    pages = "241604",
    year = "2019"
}

@article{Eckner:2024pqt,
    author = "Eckner, Christopher and Figueroa, Felipe and Tourkine, Piotr",
    title = "{On the number of Regge trajectories for dual amplitudes}",
    eprint = "2405.21057",
    archivePrefix = "arXiv",
    primaryClass = "hep-th",
    reportNumber = "LAPTH-030/24",
    doi = "10.1007/JHEP02(2025)103",
    journal = "JHEP",
    volume = "02",
    pages = "103",
    year = "2025"
}

@article{Saha:2026ftv,
    author = "Saha, Arnab Priya and Sinha, Aninda",
    title = "{Five-point partial waves, splitting constraints and hidden zeros}",
    eprint = "2601.15088",
    archivePrefix = "arXiv",
    primaryClass = "hep-th",
    month = "1",
    year = "2026"
}

@article{EliasMiro:2023fqi,
    author = "{J. Elias Mir\'{o}, A. Guerrieri, and M. A. G\"{u}m\"{u}\c{s}}",
    title = "{Extremal Higgs couplings}",
    eprint = "2311.09283",
    archivePrefix = "arXiv",
    primaryClass = "hep-ph",
    doi = "10.1103/PhysRevD.110.016007",
    journal = "Phys. Rev. D",
    volume = "110",
    pages = "016007",
    year = "2024"
}

@article{EliasMiro:2026kww,
    author = "{J. Elias Mir\'{o}, A. Guerrieri, and M. A. G\"{u}m\"{u}\c{s}}",
    title = "{The Phases of the Scalar S-Matrix Island}",
    eprint = "2605.06613",
    archivePrefix = "arXiv",
    primaryClass = "hep-th",
    month = "5",
    year = "2026"
}

@article{Guerrieri:2020bto,
    author = "Guerrieri, Andrea L. and Penedones, Joao and Vieira, Pedro",
    title = "{S-matrix bootstrap for effective field theories: massless pions}",
    eprint = "2011.02802",
    archivePrefix = "arXiv",
    primaryClass = "hep-th",
    doi = "10.1007/JHEP06(2021)088",
    journal = "JHEP",
    volume = "06",
    pages = "088",
    year = "2021"
}

@article{Wess:1971yu,
    author = "Wess, J. and Zumino, B.",
    title = "{Consequences of anomalous Ward identities}",
    doi = "10.1016/0370-2693(71)90582-X",
    journal = "Phys. Lett. B",
    volume = "37",
    pages = "95",
    year = "1971"
}

@article{Witten:1983tw,
    author = "Witten, Edward",
    title = "{Global Aspects of Current Algebra}",
    reportNumber = "PRINT-83-0262 (PRINCETON)",
    doi = "10.1016/0550-3213(83)90063-9",
    journal = "Nucl. Phys. B",
    volume = "223",
    pages = "422",
    year = "1983"
}

@article{Albert:2022oes,
    author = "Albert, Jan and Rastelli, Leonardo",
    title = "{Bootstrapping pions at large N}",
    eprint = "2203.11950",
    archivePrefix = "arXiv",
    primaryClass = "hep-th",
    reportNumber = "YITP-SB-2022-07",
    doi = "10.1007/JHEP08(2022)151",
    journal = "JHEP",
    volume = "08",
    pages = "151",
    year = "2022"
}

@article{Albert:2023jtd,
    author = "Albert, Jan and Rastelli, Leonardo",
    title = "{Bootstrapping pions at large N. Part II. Background gauge fields and the chiral anomaly}",
    eprint = "2307.01246",
    archivePrefix = "arXiv",
    primaryClass = "hep-th",
    reportNumber = "YITP-SB-2023-15",
    doi = "10.1007/JHEP09(2024)039",
    journal = "JHEP",
    volume = "09",
    pages = "039",
    year = "2024"
}

@article{Ma:2023vgc,
    author = "Ma, Teng and Pomarol, Alex and Sciotti, Francesco",
    title = "{Bootstrapping the chiral anomaly at large $N_{c}$}",
    eprint = "2307.04729",
    archivePrefix = "arXiv",
    primaryClass = "hep-th",
    doi = "10.1007/JHEP11(2023)176",
    journal = "JHEP",
    volume = "11",
    pages = "176",
    year = "2023"
}

@article{Albert:2023seb,
    author = "Albert, Jan and Henriksson, Johan and Rastelli, Leonardo and Vichi, Alessandro",
    title = "{Bootstrapping mesons at large N: Regge trajectory from spin-two maximization}",
    eprint = "2312.15013",
    archivePrefix = "arXiv",
    primaryClass = "hep-th",
    reportNumber = "YITP-SB-2023-41",
    doi = "10.1007/JHEP09(2024)172",
    journal = "JHEP",
    volume = "09",
    pages = "172",
    year = "2024"
}

@article{Dong:2024omo,
    author = "Dong, Zi-Yu and Ma, Teng and Pomarol, Alex and Sciotti, Francesco",
    title = "{Bootstrapping the chiral-gravitational anomaly}",
    eprint = "2411.14422",
    archivePrefix = "arXiv",
    primaryClass = "hep-th",
    doi = "10.1007/JHEP05(2025)114",
    journal = "JHEP",
    volume = "05",
    pages = "114",
    year = "2025"
}

@article{Dong:2025dpy,
    author = "Dong, Ziyu and Jeong, Jaehoon and Pomarol, Alex",
    title = "{Causal bounds on EFTs with anomalies with a pseudoscalar, photons, and gravitons}",
    eprint = "2510.12138",
    archivePrefix = "arXiv",
    primaryClass = "hep-th",
    reportNumber = "KIAS-Q25016",
    doi = "10.1007/JHEP02(2026)102",
    journal = "JHEP",
    volume = "02",
    pages = "102",
    year = "2026"
}

@article{Adams:2006sv,
    author = "Adams, Allan and Arkani-Hamed, Nima and Dubovsky, Sergei and Nicolis, Alberto and Rattazzi, Riccardo",
    title = "{Causality, analyticity and an IR obstruction to UV completion}",
    eprint = "hep-th/0602178",
    archivePrefix = "arXiv",
    reportNumber = "CERN-PH-TH-2006-033, HUTP-06-A0005",
    doi = "10.1088/1126-6708/2006/10/014",
    journal = "JHEP",
    volume = "10",
    pages = "014",
    year = "2006"
}

@article{Li:2023qzs,
    author = "Li, Yue-Zhou",
    title = "{Effective field theory bootstrap, large-N {\ensuremath{\chi}}PT and holographic QCD}",
    eprint = "2310.09698",
    archivePrefix = "arXiv",
    primaryClass = "hep-th",
    doi = "10.1007/JHEP01(2024)072",
    journal = "JHEP",
    volume = "01",
    pages = "072",
    year = "2024"
}

@article{Bellazzini:2020cot,
    author = "Bellazzini, Brando and Elias Mir\'o, Joan and Rattazzi, Riccardo and Riembau, Marc and Riva, Francesco",
    title = "{Positive moments for scattering amplitudes}",
    eprint = "2011.00037",
    archivePrefix = "arXiv",
    primaryClass = "hep-th",
    doi = "10.1103/PhysRevD.104.036006",
    journal = "Phys. Rev. D",
    volume = "104",
    pages = "036006",
    year = "2021"
}

@article{Caron-Huot:2020cmc,
    author = "Caron-Huot, Simon and Van Duong, Vincent",
    title = "{Extremal Effective Field Theories}",
    eprint = "2011.02957",
    archivePrefix = "arXiv",
    primaryClass = "hep-th",
    doi = "10.1007/JHEP05(2021)280",
    journal = "JHEP",
    volume = "05",
    pages = "280",
    year = "2021"
}

@article{completeness,
    author = "Calisto, Francesco and Cheung, Clifford and Remmen, Grant N. and Sciotti, Francesco and Tarquini, Michele",
    title = "{Completeness from Gravitational Scattering}",
    eprint = "2512.11955",
    archivePrefix = "arXiv",
    primaryClass = "hep-th",
    reportNumber = "CALT-TH 2025-040",
    month = "12",
    year = "2025"
}

@book{Collins:1977jy,
    author = "Collins, P. D. B.",
    title = "{An Introduction to Regge Theory and High Energy Physics}",
    doi = "10.1017/9781009403269",
    publisher = "Cambridge University Press",
    year = "1977"
}

@article{Jacob:1959at,
    author = "Jacob, M. and Wick, G. C.",
    title = "{On the General Theory of Collisions for Particles with Spin}",
    doi = "10.1006/aphy.2000.6022",
    journal = "Annals Phys.",
    volume = "7",
    pages = "404",
    year = "1959"
}

@article{SFAN,
    author = "Cheung, Clifford and Remmen, Grant N. and Sciotti, Francesco and Tarquini, Michele",
    title = "{Strings from Almost Nothing}",
    eprint = "2508.09246",
    archivePrefix = "arXiv",
    primaryClass = "hep-th",
    reportNumber = "CALT-TH 2025-027",
    month = "8",
    year = "2025"
}

@article{Pham:1985cr,
      author         = "Pham, T. N. and Truong, Tran N.",
      title          = "{Evaluation of the derivative quartic terms of the meson
                        chiral Lagrangian from forward dispersion relations}",
      journal        = "Phys. Rev.",
      volume         = "D31",
      year           = "1985",
      pages          = "3027",
      doi            = "10.1103/PhysRevD.31.3027",
      reportNumber   = "Print-85-0588 (ECOLE POLY)",
      SLACcitation   = "%%CITATION = PHRVA,D31,3027;%%"
}

@article{Ananthanarayan:1994hf,
      author         = "Ananthanarayan, B. and Toublan, D. and Wanders, G.",
      title          = "{Consistency of the chiral pion-pion scattering
                        amplitudes with axiomatic constraints}",
      journal        = "Phys. Rev.",
      volume         = "D51",
      year           = "1995",
      pages          = "1093",
      doi            = "10.1103/PhysRevD.51.1093",
      eprint         = "hep-ph/9410302",
      archivePrefix  = "arXiv",
      primaryClass   = "hep-ph",
      reportNumber   = "UNIL-TP-4-94",
      SLACcitation   = "%%CITATION = HEP-PH/9410302;%%"
}

@article{Pennington:1994kc,
      author         = "Pennington, M. R. and Portoles, J.",
      title          = "{The chiral lagrangian parameters, $\ell_1$, $\ell_2$, are determined
                        by the $\rho$-resonance}",
      journal        = "Phys. Lett.",
      volume         = "B344",
      year           = "1995",
      pages          = "399",
      doi            = "10.1016/0370-2693(94)01551-M",
      eprint         = "hep-ph/9409426",
      archivePrefix  = "arXiv",
      primaryClass   = "hep-ph",
      reportNumber   = "DTP-94-54",
      SLACcitation   = "%%CITATION = HEP-PH/9409426;%%"
}

@article{Fernandez:2022kzi,
    author = "Fernandez, Clara and Pomarol, Alex and Riva, Francesco and Sciotti, Francesco",
    title = "{Cornering large-$N_{c}$ QCD with positivity bounds}",
    eprint = "2211.12488",
    archivePrefix = "arXiv",
    primaryClass = "hep-th",
    doi = "10.1007/JHEP06(2023)094",
    journal = "JHEP",
    volume = "06",
    pages = "094",
    year = "2023"
}

@article{Jenkins:2006ia,
      author         = "Jenkins, Alejandro and O'Connell, Donal",
      title          = "{The Story of ${\cal O}$: Positivity constraints in effective
                        field theories}",
      year           = "2006",
      eprint         = "hep-th/0609159",
      archivePrefix  = "arXiv",
      primaryClass   = "hep-th",
      reportNumber   = "CALT-68-2607, MIT-CTP-3764",
      SLACcitation   = "%%CITATION = HEP-TH/0609159;%%"
}

@article{Dvali:2012zc,
      author         = "Dvali, Gia and Franca, Andre and Gomez, Cesar",
      title          = "{Road Signs for UV-Completion}",
      year           = "2012",
      eprint         = "1204.6388",
      archivePrefix  = "arXiv",
      primaryClass   = "hep-th",
      SLACcitation   = "%%CITATION = ARXIV:1204.6388;%%"
}

@article{Arkani-Hamed:2020blm,
    author = "Arkani-Hamed, Nima and Huang, Tzu-Chen and Huang, Yu-tin",
    title = "{The EFT-Hedron}",
    eprint = "2012.15849",
    archivePrefix = "arXiv",
    primaryClass = "hep-th",
    reportNumber = "NCTS-TH/2014, CALT-TH 2020-061",
    doi = "10.1007/JHEP05(2021)259",
    journal = "JHEP",
    volume = "05",
    pages = "259",
    year = "2021"
}

@article{EliasMiro:2022xaa,
    author = "Elias Miro, Joan and Guerrieri, Andrea and Gumus, Mehmet Asim",
    title = "{Bridging positivity and S-matrix bootstrap bounds}",
    eprint = "2210.01502",
    archivePrefix = "arXiv",
    primaryClass = "hep-th",
    doi = "10.1007/JHEP05(2023)001",
    journal = "JHEP",
    volume = "05",
    pages = "001",
    year = "2023"
}

@article{Guerrieri:2021ivu,
    author = "Guerrieri, Andrea and Penedones, Joao and Vieira, Pedro",
    title = "{Where Is String Theory in the Space of Scattering Amplitudes?}",
    eprint = "2102.02847",
    archivePrefix = "arXiv",
    primaryClass = "hep-th",
    doi = "10.1103/PhysRevLett.127.081601",
    journal = "Phys. Rev. Lett.",
    volume = "127",
    pages = "081601",
    year = "2021"
}

@article{Caron-Huot:2021rmr,
    author = "Caron-Huot, Simon and Mazac, Dalimil and Rastelli, Leonardo and Simmons-Duffin, David",
    title = "{Sharp boundaries for the swampland}",
    eprint = "2102.08951",
    archivePrefix = "arXiv",
    primaryClass = "hep-th",
    doi = "10.1007/JHEP07(2021)110",
    journal = "JHEP",
    volume = "07",
    pages = "110",
    year = "2021"
}

@article{Nicolis:2009qm,
    author = "Nicolis, Alberto and Rattazzi, Riccardo and Trincherini, Enrico",
    title = "{Energy's and amplitudes' positivity}",
    eprint = "0912.4258",
    archivePrefix = "arXiv",
    primaryClass = "hep-th",
    doi = "10.1007/JHEP05(2010)095",
    journal = "JHEP",
    volume = "05",
    pages = "095",
    year = "2010",
    note           = "[Erratum: \href{https://doi.org/10.1007/JHEP11(2011)128}{{\it JHEP} {\bf 11} (2011) 128}]"
}

@article{Tolley:2020gtv,
    author = "Tolley, Andrew J. and Wang, Zi-Yue and Zhou, Shuang-Yong",
    title = "{New positivity bounds from full crossing symmetry}",
    eprint = "2011.02400",
    archivePrefix = "arXiv",
    primaryClass = "hep-th",
    doi = "10.1007/JHEP05(2021)255",
    journal = "JHEP",
    volume = "05",
    pages = "255",
    year = "2021"
}

@article{Bellazzini:2025shd,
    author = "Bellazzini, Brando and Pomarol, Alex and Romano, Marcello and Sciotti, Francesco",
    title = "{(Super) gravity from positivity}",
    eprint = "2507.12535",
    archivePrefix = "arXiv",
    primaryClass = "hep-th",
    doi = "10.1007/JHEP03(2026)028",
    journal = "JHEP",
    volume = "03",
    pages = "028",
    year = "2026"
}

@article{Bellazzini:2015cra,
    author = "Bellazzini, Brando and Cheung, Clifford and Remmen, Grant N.",
    title = "{Quantum Gravity Constraints from Unitarity and Analyticity}",
    eprint = "1509.00851",
    archivePrefix = "arXiv",
    primaryClass = "hep-th",
    reportNumber = "CALT-TH-2015-044, SACLAY-T15-161",
    doi = "10.1103/PhysRevD.93.064076",
    journal = "Phys. Rev. D",
    volume = "93",
    pages = "064076",
    year = "2016"
}

@article{Cheung:2016wjt,
    author = "Cheung, Clifford and Remmen, Grant N.",
    title = "{Positivity of Curvature-Squared Corrections in Gravity}",
    eprint = "1608.02942",
    archivePrefix = "arXiv",
    primaryClass = "hep-th",
    reportNumber = "CALT-TH-2016-018",
    doi = "10.1103/PhysRevLett.118.051601",
    journal = "Phys. Rev. Lett.",
    volume = "118",
    pages = "051601",
    year = "2017"
}

@article{Camanho:2014apa,
    author = "Camanho, Xian O. and Edelstein, Jose D. and Maldacena, Juan and Zhiboedov, Alexander",
    title = "{Causality Constraints on Corrections to the Graviton Three-Point Coupling}",
    eprint = "1407.5597",
    archivePrefix = "arXiv",
    primaryClass = "hep-th",
    doi = "10.1007/JHEP02(2016)020",
    journal = "JHEP",
    volume = "02",
    pages = "020",
    year = "2016"
}

@article{Gruzinov:2006ie,
    author = "Gruzinov, A. and Kleban, M.",
    title = "{Causality Constrains Higher Curvature Corrections to Gravity}",
    eprint = "hep-th/0612015",
    archivePrefix = "arXiv",
    doi = "10.1088/0264-9381/24/13/N02",
    journal = "Class. Quant. Grav.",
    volume = "24",
    pages = "3521",
    year = "2007"
}

@article{Arkani-Hamed:2021ajd,
    author = "Arkani-Hamed, Nima and Huang, Yu-tin and Liu, Jin-Yu and Remmen, Grant N.",
    title = "{Causality, unitarity, and the weak gravity conjecture}",
    eprint = "2109.13937",
    archivePrefix = "arXiv",
    primaryClass = "hep-th",
    doi = "10.1007/JHEP03(2022)083",
    journal = "JHEP",
    volume = "03",
    pages = "083",
    year = "2022"
}

@article{Cheung:2018cwt,
    author = "Cheung, Clifford and Liu, Junyu and Remmen, Grant N.",
    title = "{Proof of the Weak Gravity Conjecture from Black Hole Entropy}",
    eprint = "1801.08546",
    archivePrefix = "arXiv",
    primaryClass = "hep-th",
    reportNumber = "CALT-TH-2018-007",
    doi = "10.1007/JHEP10(2018)004",
    journal = "JHEP",
    volume = "10",
    pages = "004",
    year = "2018"
}

@article{Cheung:2019cwi,
    author = "Cheung, Clifford and Liu, Junyu and Remmen, Grant N.",
    title = "{Entropy Bounds on Effective Field Theory from Rotating Dyonic Black Holes}",
    eprint = "1903.09156",
    archivePrefix = "arXiv",
    primaryClass = "hep-th",
    reportNumber = "CALT-TH-2019-003",
    doi = "10.1103/PhysRevD.100.046003",
    journal = "Phys. Rev. D",
    volume = "100",
    pages = "046003",
    year = "2019"
}

@article{Cheung:2014ega,
      author         = "Cheung, Clifford and Remmen, Grant N.",
      title          = "{Infrared Consistency and the Weak Gravity Conjecture}",
      journal        = "JHEP",
      volume         = "12",
      year           = "2014",
      pages          = "087",
      doi            = "10.1007/JHEP12(2014)087",
      eprint         = "1407.7865",
      archivePrefix  = "arXiv",
      primaryClass   = "hep-th",
      reportNumber   = "CALT-TH-2014-146",
      SLACcitation   = "%%CITATION = ARXIV:1407.7865;%%"
}

@article{Bellazzini:2019xts,
    author = "Bellazzini, Brando and Lewandowski, Matthew and Serra, Javi",
    title = "{Positivity of Amplitudes, Weak Gravity Conjecture, and Modified Gravity}",
    eprint = "1902.03250",
    archivePrefix = "arXiv",
    primaryClass = "hep-th",
    doi = "10.1103/PhysRevLett.123.251103",
    journal = "Phys. Rev. Lett.",
    volume = "123",
    pages = "251103",
    year = "2019"
}

@article{Brower:1974yv,
    author = "Brower, R. C. and DeTar, Carleton E. and Weis, J. H.",
    title = "{Regge Theory for Multiparticle Amplitudes}",
    reportNumber = "MIT-CTP-395, CERN-TH-1817",
    doi = "10.1016/0370-1573(74)90012-X",
    journal = "Phys. Rept.",
    volume = "14",
    pages = "257",
    year = "1974"
}

@article{Caron-Huot:2022ugt,
    author = "Caron-Huot, Simon and Li, Yue-Zhou and Parra-Martinez, Julio and Simmons-Duffin, David",
    title = "{Causality constraints on corrections to Einstein gravity}",
    eprint = "2201.06602",
    archivePrefix = "arXiv",
    primaryClass = "hep-th",
    reportNumber = "CALT-TH 2021-003",
    doi = "10.1007/JHEP05(2023)122",
    journal = "JHEP",
    volume = "05",
    pages = "122",
    year = "2023"
}

@article{Caron-Huot:2022jli,
    author = "Caron-Huot, Simon and Li, Yue-Zhou and Parra-Martinez, Julio and Simmons-Duffin, David",
    title = "{Graviton partial waves and causality in higher dimensions}",
    eprint = "2205.01495",
    archivePrefix = "arXiv",
    primaryClass = "hep-th",
    reportNumber = "CALT-TH 2022-17",
    doi = "10.1103/PhysRevD.108.026007",
    journal = "Phys. Rev. D",
    volume = "108",
    pages = "026007",
    year = "2023"
}

@article{Cheung:2016yqr,
    author = "Cheung, Clifford and Remmen, Grant N.",
    title = "{Positive Signs in Massive Gravity}",
    eprint = "1601.04068",
    archivePrefix = "arXiv",
    primaryClass = "hep-th",
    reportNumber = "CALT-TH-2015-062",
    doi = "10.1007/JHEP04(2016)002",
    journal = "JHEP",
    volume = "04",
    pages = "002",
    year = "2016"
}

@article{deRham:2017xox,
      author         = "de Rham, Claudia and Melville, Scott and Tolley, Andrew
                        J.",
      title          = "{Improved Positivity Bounds and Massive Gravity}",
      journal        = "JHEP",
      volume         = "04",
      year           = "2018",
      pages          = "083",
      doi            = "10.1007/JHEP04(2018)083",
      eprint         = "1710.09611",
      archivePrefix  = "arXiv",
      primaryClass   = "hep-th",
      reportNumber   = "IMPERIAL-TP-2017-AT-04",
      SLACcitation   = "%%CITATION = ARXIV:1710.09611;%%"
}

@article{Bellazzini:2023nqj,
    author = "Bellazzini, Brando and Isabella, Giulia and Ricossa, Sergio and Riva, Francesco",
    title = "{Massive gravity is not positive}",
    eprint = "2304.02550",
    archivePrefix = "arXiv",
    primaryClass = "hep-th",
    doi = "10.1103/PhysRevD.109.024051",
    journal = "Phys. Rev. D",
    volume = "109",
    pages = "024051",
    year = "2024"
}

@article{Bern:2021ppb,
    author = "Bern, Zvi and Kosmopoulos, Dimitrios and Zhiboedov, Alexander",
    title = "{Gravitational effective field theory islands, low-spin dominance, and the four-graviton amplitude}",
    eprint = "2103.12728",
    archivePrefix = "arXiv",
    primaryClass = "hep-th",
    reportNumber = "CERN-TH-2021-035",
    doi = "10.1088/1751-8121/ac0e51",
    journal = "J. Phys. A",
    volume = "54",
    pages = "344002",
    year = "2021"
}

@article{Berman:2023jys,
    author = "Berman, Justin and Elvang, Henriette and Herderschee, Aidan",
    title = "{Flattening of the EFT-hedron: supersymmetric positivity bounds and the search for string theory}",
    eprint = "2310.10729",
    archivePrefix = "arXiv",
    primaryClass = "hep-th",
    doi = "10.1007/JHEP03(2024)021",
    journal = "JHEP",
    volume = "03",
    pages = "021",
    year = "2024"
}

@article{Berman:2024wyt,
    author = "Berman, Justin and Elvang, Henriette",
    title = "{Corners and islands in the S-matrix bootstrap of the open superstring}",
    eprint = "2406.03543",
    archivePrefix = "arXiv",
    primaryClass = "hep-th",
    reportNumber = "LCTP-24-10",
    doi = "10.1007/JHEP09(2024)076",
    journal = "JHEP",
    volume = "09",
    pages = "076",
    year = "2024"
}

@article{Albert:2024yap,
    author = "Albert, Jan and Knop, Waltraut and Rastelli, Leonardo",
    title = "{Where is tree-level string theory?}",
    eprint = "2406.12959",
    archivePrefix = "arXiv",
    primaryClass = "hep-th",
    reportNumber = "YITP-SB-2024-12",
    doi = "10.1007/JHEP02(2025)157",
    journal = "JHEP",
    volume = "02",
    pages = "157",
    year = "2025"
}

@article{Berman:2024eid,
    author = "Berman, Justin and Elvang, Henriette and Geiser, Nicholas and Lin, Loki L.",
    title = "{Bootstrapping Extremal Scalar Amplitudes With and Without Supersymmetry}",
    eprint = "2412.13368",
    archivePrefix = "arXiv",
    primaryClass = "hep-th",
    reportNumber = "LCTP-24-23",
    month = "12",
    year = "2024"
}

@article{Berman:2025owb,
    author = "Berman, Justin and Elvang, Henriette and Figueiredo, Carolina",
    title = "{Splitting regions and shrinking islands from higher point constraints}",
    eprint = "2506.22538",
    archivePrefix = "arXiv",
    primaryClass = "hep-th",
    reportNumber = "LITP-25-08",
    doi = "10.1007/JHEP10(2025)226",
    journal = "JHEP",
    volume = "10",
    pages = "226",
    year = "2025"
}

@article{Elvang:2026pmc,
    author = "Elvang, Henriette and Herderschee, Aidan and Morales, Roger",
    title = "{String Theory from Maximal Supersymmetry}",
    eprint = "2601.11705",
    archivePrefix = "arXiv",
    primaryClass = "hep-th",
    reportNumber = "LITP-25-18",
    month = "1",
    year = "2026"
}

@article{Remmen:2021zmc,
    author = "Remmen, Grant N.",
    title = "{Amplitudes and the Riemann Zeta Function}",
    eprint = "2108.07820",
    archivePrefix = "arXiv",
    primaryClass = "hep-th",
    doi = "10.1103/PhysRevLett.127.241602",
    journal = "Phys. Rev. Lett.",
    volume = "127",
    pages = "241602",
    year = "2021"
}

@article{Huang:2022mdb,
    author = "Huang, Yu-tin and Remmen, Grant N.",
    title = "{UV-complete gravity amplitudes and the triple product}",
    eprint = "2203.00696",
    archivePrefix = "arXiv",
    primaryClass = "hep-th",
    doi = "10.1103/PhysRevD.106.L021902",
    journal = "Phys. Rev. D",
    volume = "106",
    pages = "L021902",
    year = "2022"
}

@article{Cheung:2022mkw,
    author = "Cheung, Clifford and Remmen, Grant N.",
    title = "{Veneziano variations: how unique are string amplitudes?}",
    eprint = "2210.12163",
    archivePrefix = "arXiv",
    primaryClass = "hep-th",
    doi = "10.1007/JHEP01(2023)122",
    journal = "JHEP",
    volume = "01",
    pages = "122",
    year = "2023"
}

@article{Cheung:2023adk,
    author = "Cheung, Clifford and Remmen, Grant N.",
    title = "{Stringy dynamics from an amplitudes bootstrap}",
    eprint = "2302.12263",
    archivePrefix = "arXiv",
    primaryClass = "hep-th",
    reportNumber = "CALT-TH 2023-006",
    doi = "10.1103/PhysRevD.108.026011",
    journal = "Phys. Rev. D",
    volume = "108",
    pages = "026011",
    year = "2023"
}

@article{Cheung:2023uwn,
    author = "Cheung, Clifford and Remmen, Grant N.",
    title = "{Bespoke dual resonance}",
    eprint = "2308.03833",
    archivePrefix = "arXiv",
    primaryClass = "hep-th",
    reportNumber = "CALT-TH 2023-026",
    doi = "10.1103/PhysRevD.108.086009",
    journal = "Phys. Rev. D",
    volume = "108",
    pages = "086009",
    year = "2023"
}

@article{Cheung:2024uhn,
    author = "Cheung, Clifford and Hillman, Aaron and Remmen, Grant N.",
    title = "{Bootstrap Principle for the Spectrum and Scattering of Strings}",
    eprint = "2406.02665",
    archivePrefix = "arXiv",
    primaryClass = "hep-th",
    reportNumber = "CALT-TH 2024-022",
    doi = "10.1103/PhysRevLett.133.251601",
    journal = "Phys. Rev. Lett.",
    volume = "133",
    pages = "251601",
    year = "2024"
}

@article{Cheung:2024obl,
    author = "Cheung, Clifford and Hillman, Aaron and Remmen, Grant N.",
    title = "{Uniqueness criteria for the Virasoro-Shapiro amplitude}",
    eprint = "2408.03362",
    archivePrefix = "arXiv",
    primaryClass = "hep-th",
    reportNumber = "CALT-TH 2024-030",
    doi = "10.1103/PhysRevD.111.086034",
    journal = "Phys. Rev. D",
    volume = "111",
    pages = "086034",
    year = "2025"
}

@article{Eckner:2024ggx,
    author = "Eckner, Christopher and Figueroa, Felipe and Tourkine, Piotr",
    title = "{Regge bootstrap: From linear to nonlinear trajectories}",
    eprint = "2401.08736",
    archivePrefix = "arXiv",
    primaryClass = "hep-th",
    doi = "10.1103/PhysRevD.111.126005",
    journal = "Phys. Rev. D",
    volume = "111",
    pages = "126005",
    year = "2025"
}

@article{Haring:2023zwu,
    author = {H{\"a}ring, Kelian and Zhiboedov, Alexander},
    title = "{The stringy S-matrix bootstrap: maximal spin and superpolynomial softness}",
    eprint = "2311.13631",
    archivePrefix = "arXiv",
    primaryClass = "hep-th",
    reportNumber = "CERN-TH-2023-214",
    doi = "10.1007/JHEP10(2024)075",
    journal = "JHEP",
    volume = "10",
    pages = "075",
    year = "2024"
}

@article{Haring:2024wyz,
    author = {H{\"a}ring, Kelian and Zhiboedov, Alexander},
    title = "{What is the graviton pole made of?}",
    eprint = "2410.21499",
    archivePrefix = "arXiv",
    primaryClass = "hep-th",
    reportNumber = "CERN-TH-2024-178",
    month = "10",
    year = "2024"
}

@article{Arkani-Hamed:2022gsa,
    author = "Arkani-Hamed, Nima and Eberhardt, Lorenz and Huang, Yu-tin and Mizera, Sebastian",
    title = "{On unitarity of tree-level string amplitudes}",
    eprint = "2201.11575",
    archivePrefix = "arXiv",
    primaryClass = "hep-th",
    doi = "10.1007/JHEP02(2022)197",
    journal = "JHEP",
    volume = "02",
    pages = "197",
    year = "2022"
}

@article{Bhardwaj:2024klc,
    author = "Bhardwaj, Rishabh and Spradlin, Marcus and Volovich, Anastasia and Weng, He-Chen",
    title = "{Unitarity of bespoke amplitudes}",
    eprint = "2406.04410",
    archivePrefix = "arXiv",
    primaryClass = "hep-th",
    doi = "10.1103/PhysRevD.110.106016",
    journal = "Phys. Rev. D",
    volume = "110",
    pages = "106016",
    year = "2024"
}

@article{Freytsis:2022aho,
    author = "Freytsis, Marat and Kumar, Soubhik and Remmen, Grant N. and Rodd, Nicholas L.",
    title = "{Multifield positivity bounds for inflation}",
    eprint = "2210.10791",
    archivePrefix = "arXiv",
    primaryClass = "hep-th",
    reportNumber = "CERN-TH-2022-160",
    doi = "10.1007/JHEP09(2023)041",
    journal = "JHEP",
    volume = "09",
    pages = "041",
    year = "2023"
}

@article{Remmen:2019cyz,
    author = "Remmen, Grant N. and Rodd, Nicholas L.",
    title = "{Consistency of the Standard Model Effective Field Theory}",
    eprint = "1908.09845",
    archivePrefix = "arXiv",
    primaryClass = "hep-ph",
    doi = "10.1007/JHEP12(2019)032",
    journal = "JHEP",
    volume = "12",
    pages = "032",
    year = "2019"
}

@article{Remmen:2020vts,
    author = "Remmen, Grant N. and Rodd, Nicholas L.",
    title = "{Flavor Constraints from Unitarity and Analyticity}",
    eprint = "2004.02885",
    archivePrefix = "arXiv",
    primaryClass = "hep-ph",
    doi = "10.1103/PhysRevLett.127.149901",
    journal = "Phys. Rev. Lett.",
    volume = "125",
    pages = "081601",
    year = "2020",   
    note           = "[Erratum: \href{https://doi.org/10.1103/PhysRevLett.127.149901}{{\it Phys. Rev. Lett.} {\bf 127}, 149901 (2021)}]"
}

@article{Remmen:2020uze,
    author = "Remmen, Grant N. and Rodd, Nicholas L.",
    title = "{Signs, spin, SMEFT: Sum rules at dimension six}",
    eprint = "2010.04723",
    archivePrefix = "arXiv",
    primaryClass = "hep-ph",
    doi = "10.1103/PhysRevD.105.036006",
    journal = "Phys. Rev. D",
    volume = "105",
    pages = "036006",
    year = "2022"
}

@article{Bellazzini:2016xrt,
    author = "Bellazzini, Brando",
    title = "{Softness and amplitudes' positivity for spinning particles}",
    eprint = "1605.06111",
    archivePrefix = "arXiv",
    primaryClass = "hep-th",
    reportNumber = "SACLAY-T16-038",
    doi = "10.1007/JHEP02(2017)034",
    journal = "JHEP",
    volume = "02",
    pages = "034",
    year = "2017"
}

@article{Remmen:2022orj,
    author = "Remmen, Grant N. and Rodd, Nicholas L.",
    title = "{Spinning sum rules for the dimension-six SMEFT}",
    eprint = "2206.13524",
    archivePrefix = "arXiv",
    primaryClass = "hep-ph",
    reportNumber = "CERN-TH-2022-105",
    doi = "10.1007/JHEP09(2022)030",
    journal = "JHEP",
    volume = "09",
    pages = "030",
    year = "2022"
}

@article{Remmen:2024hry,
    author = "Remmen, Grant N. and Rodd, Nicholas L.",
    title = "{Positively identifying Higgs effective field theory or standard model effective field theory}",
    eprint = "2412.07827",
    archivePrefix = "arXiv",
    primaryClass = "hep-ph",
    doi = "10.1103/vj7j-zj11",
    journal = "Phys. Rev. D",
    volume = "113",
    pages = "036027",
    year = "2026"
}

@article{Bi:2019phv,
    author = "Bi, Qi and Zhang, Cen and Zhou, Shuang-Yong",
    title = "{Positivity constraints on aQGC: carving out the physical parameter space}",
    eprint = "1902.08977",
    archivePrefix = "arXiv",
    primaryClass = "hep-ph",
    reportNumber = "USTC-ICTS-19-01",
    doi = "10.1007/JHEP06(2019)137",
    journal = "JHEP",
    volume = "06",
    pages = "137",
    year = "2019"
}

@article{Zhang:2018shp,
    author = "Zhang, Cen and Zhou, Shuang-Yong",
    title = "{Positivity bounds on vector boson scattering at the LHC}",
    eprint = "1808.00010",
    archivePrefix = "arXiv",
    primaryClass = "hep-ph",
    reportNumber = "USTC-ICTS-18-13",
    doi = "10.1103/PhysRevD.100.095003",
    journal = "Phys. Rev. D",
    volume = "100",
    pages = "095003",
    year = "2019"
}

\medskip
\noindent {\bf Appendix}

\mysubsec{Matrix elements} We compute the matrix element $F^J_{MM'}(\vec z)$ for an angular momentum $J$. Using
$U[{\cal R}(\vec z)]=e^{-z_1J_+}e^{-z_2J_-}e^{-z_3 J_0}$ in the unitary representation,
we expand the two unipotent exponentials. Let $k$ be the number of $J_-$ actions.
Then $J_-^k$ lowers $M'$ to $M'-k$, while $e^{-z_3J_0}$ gives
$e^{-z_3M'}$. To reach the final state $M$, the remaining exponential must provide
$M-M'+k$ powers of $J_+$. Therefore,
\begin{align}
&F^J_{MM'}(\vec z)
=\sqrt{
\frac{(J+M)!(J+M')!}{(J-M)!(J-M')!}
}
\nonumber
\\
\!\times &
\sum_{k}
\frac{(-z_1)^{M-M'+k}(-z_2)^k e^{-z_3M'}}
{(M-M'+k)!\,k!}
\frac{(J-M'+k)!}{(J+M'-k)!},
\end{align}
where the summation range is fixed by the allowed magnetic quantum numbers,
$\max(0,M'-M)\leq k\leq J+M'$.

\medskip
\mysubsec{WZW leading term}As an example, we consider the Wess-Zumino-Witten (WZW) interaction for pions~\cite{Wess:1971yu,Witten:1983tw}. This term gives the
leading odd-point amplitude in the chiral Lagrangian and plays a central role
in the five-point contribution to the planar large-$N_c$ multipositivity bounds
studied in Ref.~\cite{Cheung:2026lpv}. Here we derive the spin-resolved input to the five-point dispersion relation using the WZW amplitude as an explicit example.

From the WZW action, the Lagrangian reads
\begin{align}
{\cal L}_{\rm WZW}
=
\frac{2\tilde \kappa}{15}
\,
\epsilon^{\mu\nu\rho\sigma}
\mbox{Tr}
(\pi \partial_\mu\pi 
\partial_\nu\pi 
\partial_\rho\pi 
\partial_\sigma\pi )
+\cdots ,
\end{align}
with 
$\tilde \kappa=N_c/(4\sqrt{2}F_{\pi}^5)$ fixed by the QCD chiral anomaly.
On the chart in Eq.~\eqref{eq:2-to-3_momenta}, the low-energy WZW amplitude is
\begin{align}
A^{(5)}_{\rm WZW}
=
\frac{i\tilde\kappa}{6}\,
b_1\Big[b_2s+a(s_{45}+b_1b_2)\Big].
\label{eq:low-e_WZW}
\end{align}
\vspace{0.1cm}

In large-$N_c$ QCD, the dispersion relation in the factorization invariant $s$
requires the boundary term at $s\to\infty$ to vanish with the other Mandelstams
fixed. Planar Regge behavior at tree level gives
$A^{(n)}\sim s^{\max_i[\alpha_i]}$ $(i=1,...,n-3)$, where $\alpha_i$ are the Regge intercepts
associated with the half-ladder cuts~\cite{Brower:1974yv,Collins:1977jy}. Since the leading large-$N_c$
trajectories contain no pomeron, i.e., $\alpha_i<1$, we have $A^{(5)}/s\to0$ at $s\to\infty$.

We can therefore connect the low-energy WZW coefficient to the UV resonance
data by the contour integral of $A^{(5)}/s$. To take the $s_{45}\to0$ limit in
the five-point PW in Eq.~\eqref{eq:5pt_PWs}, we combine the coefficient and the rotation matrix element of the two-particle
wavefunction inside the three-particle block through the $J_2$ sum:
\begin{align}
&
\sum_{J_2}
H^{J_1}_{|J_2M_2}(s,s_{45})\,
F^{J_2}_{M_20}\bigg(
-\frac{b_1\sqrt{s_{45}}}{s-s_{45}},
\frac{b_2}{\sqrt{s_{45}}},
0
\bigg)
\nonumber
\\
&\equiv \sum_{m,n,l}\frac{{\cal H}^{J_1}_{M_2;mnl}(s)}{m!n!l!}
\frac{b_1^m b_2^n s_{45}^l}{s^{(m+n+2l)/2}},
\label{eq:}
\end{align}
where we take the full amplitude to be analytic in the collinear limit $s_{45}\to0$ of the $(4,5)$ subsystem. In this limit, the subsystem spin tower
is reorganized into the reduced coefficient ${\cal H}^{J_1}_{M_2;mnl}$, while
$M_2$ keeps track of the $J_0$ projection.

At a pole of a resonance with spin $J_i$ and mass $m_i$, the reduced coefficient
factorizes as ${\cal H}^{J_i}_{M;mnl}\to
g^{(2\pi)}_{J_i}g^{(3\pi)}_{J_iM;mnl}$ in terms of the couplings to the two- and
three-pion blocks. Extracting the $b_1b_2$ coefficient at $a=s_{45}=0$ in Eqs.~\eqref{eq:5pt_PWs} and~\eqref{eq:low-e_WZW} gives
\begin{align}
\frac{i\tilde{\kappa}}{6}=-
\sum_{i}\frac{g^{(2\pi)}_{J_i}}{m_i^2}
\bigg\{g^{(3\pi)}_{J_i0;110}
-
\mathcal{J}_i
g^{(3\pi)}_{J_i,-1;010}
\bigg\},
\end{align}
with $\mathcal{J}_i=\sqrt{J_i(J_i+1)}$. Thus the WZW coefficient is resolved into UV resonance data with definite spin.
The first term can receive contributions from resonances of any spin, while the
spin-0 contribution is absent in the second term.

\end{document}